# An Unbiased Near-infrared Interferometric Survey for Hot Exozodiacal Dust

Steve Ertel[1,2,3]
Jean-Charles Augereau[2,3]
Olivier Absil[4]
Denis Defrère[5]
Jean-Baptiste Le Bouquin[2,3]
Lindsay Marion[4]
Amy Bonsor[6]
Jérémy Lebreton[7,8]

[1] ESO
[2] Université Grenoble Alpes, France
[3] CNRS, Institut de Planétologie et d'Astrophysique de Grenoble, France
[4] Départment d'Astrophysique, Géophysique et Océanographie, Université de Liège, Belgium
[5] Department of Astronomy, Steward Observatory, University of Arizona, USA
[6] School of Physics, H. H. Wills Physics Laboratory, University of Bristol, United Kingdom
[7] NASA Exoplanet Science Institute, California Institute of Technology, Pasadena, USA
[8] Infrared Processing and Analysis Center, California Institute of Technology, Pasadena, USA

Exozodiacal dust is warm or hot dust found in the inner regions of planetary systems orbiting main sequence stars, in or around their habitable zones. The dust can be the most luminous component of extrasolar planetary systems, but predominantly emits in the near- to mid-infrared where it is outshone by the host star. Interferometry provides a unique method of separating this dusty emission from the stellar emission. The visitor instrument PIONIER at the Very Large Telescope Interferometer (VLTI) has been used to search for hot exozodiacal dust around a large sample of nearby main sequence stars. The results of this survey are summarised: 9 out of 85 stars show excess exozodiacal emission over the stellar photospheric emission.

Planetesimals and comets are a major component of the Solar System (in the Kuiper Belt and the asteroid belt), as well as of extrasolar planetary systems, where they occur in debris discs. Besides the planets, they are the main outcome of the planet formation process. Studying the composition and distribution of the dust produced in debris discs through collisions and outgassing of these larger bodies is a powerful tool that can help to constrain the architecture, dynamics and evolution of extrasolar planetary systems. However, debris discs that are relatively easy to detect are located several astronomical units (au) to a few hundreds of au from their host stars, and thus only trace the outer regions of planetary systems. In order to study the inner regions close to the habitable zones, one has to study warm and hot dust closer to the star. This dust is called exozodiacal dust, or exozodi for short, by analogue with the Zodiacal dust in the Solar System.

The Zodiacal light can be observed on dark nights directly after dusk and before dawn as a cone of faint light stretching from the horizon in the west (after dusk) or in the east (before dawn). It is caused by sunlight scattered off small dust particles close to the orbit of the Earth. More generally, Zodiacal dust is distributed in a disc inside the asteroid belt, extending all the way down to the sublimation distance of the dust from the Sun, which corresponds to a few Solar radii. The dust temperatures range from about 100 K to about 2000 K, depending on the distance from the Sun. In the innermost regions it forms the Fraunhofer corona (F-corona) of the Sun, a region of the corona where the prominent absorption lines in the Solar spectrum are visible because the light seen there is nearly unaltered sunlight scattered by the dust particles. It is noteworthy that the Zodiacal light is the most luminous component of the Solar System after the Sun itself.

Not unlike the Zodiacal dust, exozodiacal dust is located in the inner regions of extrasolar planetary systems, within a few au of main sequence stars. This region often encompasses their habitable zone. Historically, this circumstance has brought it a lot of attention, because the presence of exozodis is expected to complicate the direct-imaging detection and characterisation of Earth-like planets in the habitable zones around other stars by future space missions. The faint light of these potential planets can be hidden in the extended emission of the dust disc. The structures created in the dust distribution due to planet–disc interaction may provide clues pointing towards the presence of a planet, but clumps in the dust distribution may also be misinterpreted as actual planets due to the limited resolution and sensitivity of current instruments. Thus, detecting and characterising exozodiacal dust systems provides critical input for the design of such space missions. However, detecting the dust itself with present instruments is complicated by the fact that the warm and hot emission peaks at near- to mid-infrared wavelengths where the dust emission is outshone by the host star. Thus, only a few very bright — and perhaps unrepresentative — systems can be detected photometrically.

## Infrared interferometric detection of exozodis

Due to the small extent of exozodiacal systems — one au at a typical distance of 10 pc for nearby stars corresponds to an angular size of 100 milliarcseconds — only interferometry is currently able to spatially resolve them. When used at baselines of a few tens of metres, near-infrared interferometry is able to fully resolve the extended emission of the dust disc while the star still remains largely unresolved. The result is a small deficit in the measured squared visibilities (the main observable of infrared interferometry) compared to the values expected from the star alone (see Figure 1 for an explanation). Using this technique the disc can be spatially disentangled from the star, allowing the disc to be detected and the ratio between the disc and stellar emission measured. This method has so far been the most powerful and efficient tool in the search for faint exozodiacal dust.

However, the dust detected by this method in the near-infrared (NIR) is expected to be very hot, close to its sublimation temperature, and its relation to slightly cooler dust in the habitable zone is unclear. This habitable zone dust is brighter in the mid-infrared (MIR) where it can be detected by the VLTI MID-infrared interferometric instrument (MIDI) for a few bright systems (e.g., Smith et al., 2009) or more efficiently by nulling interferometry. With this latter method the stellar light from two telescopes is brought to destructive interference, while light off-centre from the star is transmitted, which improves the dynamic range of the observations. Both methods have been used in parallel in the past to search for exozodiacal dust, mostly with the Fiber Linked Unit for Optical Recombination (FLUOR) instrument at the Center for High Angular Resolution Astronomy (CHARA) array (in the NIR; e.g., Absil et al., 2006) on Mt. Wilson, California and the Keck Interferometer Nuller (in the MIR; Mennesson et al., 2014) on Mauna Kea, Hawaii.

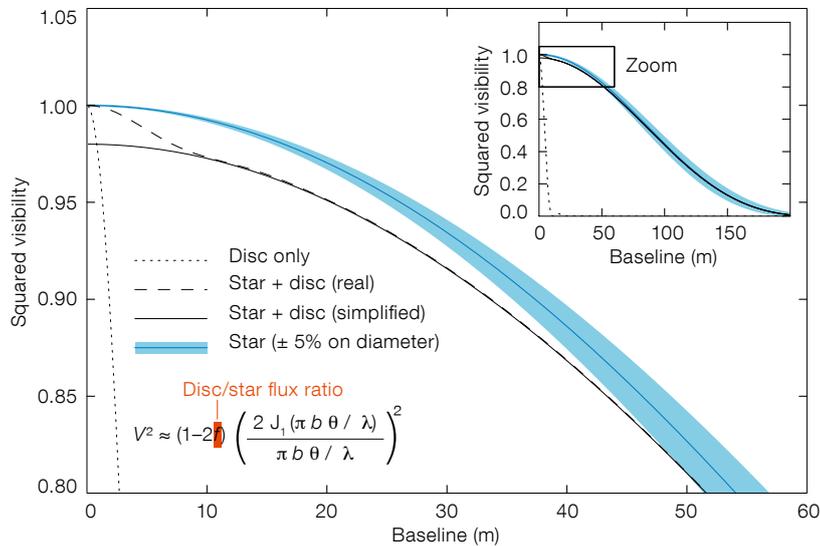

Figure 1. (Left) Illustration of the detection strategy for exozodis. The dashed curve shows a realistic case assuming a uniform disc for both the star and the flux distribution from the exozodiacal dust and a disc-to-star flux ratio of $f = 1\,\%$. For the simplified case, the solid curve shows the same assumptions, but with the approximation following the equation. Diameters of the star and (face-on) disc have been chosen to be 2.5 milliarcseconds (about an A-type star at 10 pc) and 500 milliarcseconds (5 au at 10 pc), but exact numbers are not relevant for this illustration.

Figure 2. (Below) A diagram (not to scale) to illustrate the scattering of planetesimals by an outer planet, that leads to an exchange of angular momentum and the outward migration of that planet (Bonsor et al., 2014). Some of the scattered particles are ejected, whilst some are scattered into the inner planetary system, where they interact with the inner planets. This scattering leads to a flux of material into the exozodi region.

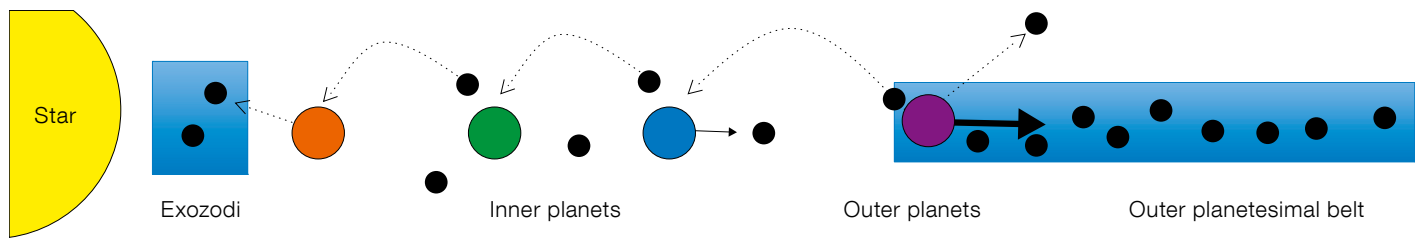

Due to the recent development of the visitor instrument PIONIER (Precision Integrated Optics Near Infrared ExpeRiment; LeBouquin et al., 2011) at the VLTI, which operates in the $H$-band, the search for exozodis in the NIR has become more efficient and of similar accuracy to the previous surveys, allowing for a significant increase in the number of surveyed stars. The sensitivity of the available instruments is, however, only capable of detecting exozodis a couple of hundred times brighter (more massive) than the Zodiacal dust. This state-of-the-art exozodi sensitivity is approximately one order of magnitude larger than that required to prepare future exoEarth imaging missions. The Large Binocular Telescope Interferometer (LBTI) is designed to achieve the required sensitivity and will soon start a MIR survey of 50 to 60 carefully chosen nearby main sequence stars (Weinberger et al., 2015).

## Potential origins of exozodiacal dust

The fact that many exozodiacal systems have already been found, given the limited sensitivity of present instruments, is surprising (Absil et al., 2013). Such high levels of dust are difficult to sustain. The Zodiacal dust has two main origins: collisions of asteroids in the asteroid belt result in dust that is dragged inwards by the interaction with stellar radiation (Poynting–Robertson drag) and the evaporation of comets heated when they get close to the Sun supplies dust. While these scenarios might work for a few MIR-detected systems, they generally do not work well for more massive systems, as have been detected in the NIR around other stars. If the dust density in the disc is too high, the dust will further collide and be destroyed before being dragged inwards to regions where it is detected. Furthermore, a higher dust mass would require a larger number of planetesimals which are colliding more often to produce the dust present; this process would in turn destroy the planetesimals faster, so that at the ages of the systems observed, few planetesimals would be left (Wyatt et al., 2007). The production of the observed amounts of dust through comet evaporation would require a large number of comets, approximately one thousand events per year similar to Hale–Bopp reaching its perihelion.

A potential scenario explaining such a large number of comets would be a connection to an outer debris disc where planetesimals are scattered inwards through gravitational interaction with an existing planetary system (see Figure 2). Considering the realistic interaction of these planets with the planetesimals (planetesimal-driven planetary migration), this scenario is able to produce detectable exozodis even if the outer debris disc were too faint to be detected with present instruments (Bonsor et al., 2014). However, such a mechanism puts strong constraints on the architecture of the underlying planetary system. While this would be a very interesting scenario, it is questionable whether all systems observed are likely to meet these constraints. A combination of planetesimal-driven migration and dust trapping could explain the presence of the detected dust. Potential dust-trapping mechanisms are suggested from the realistic treatment of the dust sublimation (Lebreton et al., 2013) or the interaction of dust grains with the stellar magnetic field (Czechowski & Mann, 2010).

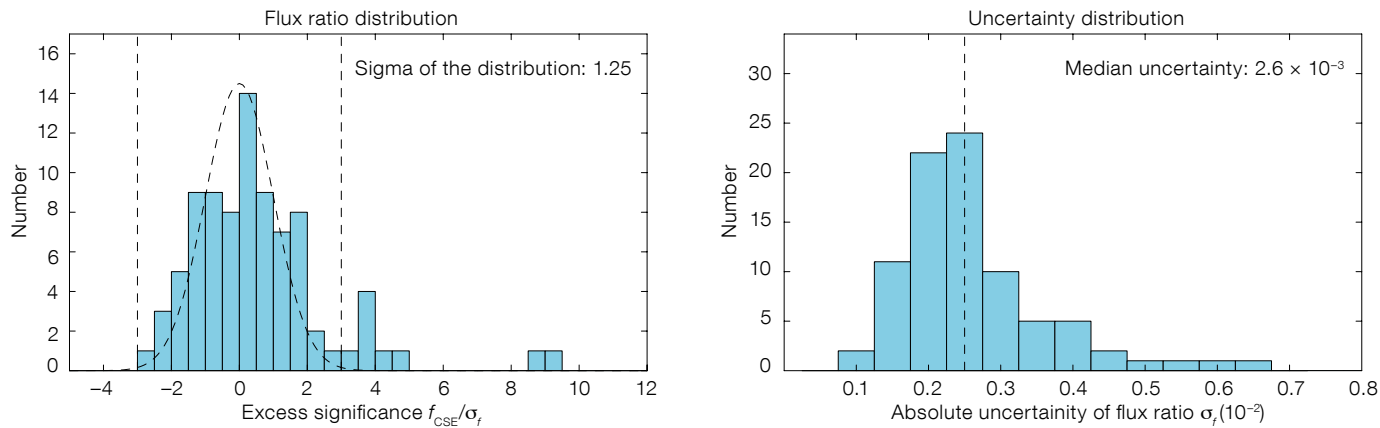

Figure 3. Excess distribution (left) and distribution of uncertainties (right) on the disc-to-star flux ratio are shown. The Gaussian overplotted on the excess distribution has a width of σ = 1 and is used to guide the eye, illustrating that the data are consistent with this ideal behaviour. Vertical dashed lines are plotted at $f = -3\sigma$ and $+3\sigma$ for the excess distribution and at the median uncertainty ($2.6 \times 10^{-3}$) for the uncertainty distribution.

### An unbiased near-infrared interferometric survey

Strong constraints on the potential origins of exozodiacal dust, described above, can be placed by a statistical analysis of the detection rate and excess levels of exozodiacal dust with respect to other properties of the systems, such as stellar spectral type, age of the system, or the presence of a detectable cool debris disc in the outer regions of the systems. A dependence on the stellar spectral type (i.e., stellar mass), similar to the one that has been found for debris discs and younger protoplanetary discs, would suggest a similar or even common origin of the material detected in these different dust populations. For these systems, more massive discs are present around more massive stars. A strong dependence on the system's age (most exozodis are detected around very young stars) would suggest that the origin is a collisionally depleting planetesimal belt. Alternatively a strong correlation with the presence of an outer disc would suggest that the hot material is supplied through some mechanism from this outer reservoir.

Our team carried out the first large, NIR interferometric survey for hot exozodiacal dust in order to statistically address the questions on the origin and evolution of these enigmatic systems (Absil et al., 2013; Ertel et al., 2014). We started our search using CHARA/FLUOR, the first instrument to routinely provide the required accuracy of the measurements. However, FLUOR can only use two telescopes at a time, which results in a limited observational efficiency. Thus, in about seven years, only around 40 stars could be observed with sufficient accuracy, but this still allowed the first statistical conclusions to be drawn (Absil et al., 2013).

The development of the PIONIER instrument for the VLTI, which can use four telescopes simultaneously, allowed a significant increase in our observing efficiency, so that in only 12 nights during 2012 a total of 92 stars could be observed. A cumulative median 1σ accuracy per target of 0.26 % on the disc-to-star flux ratio was reached (Figure 3). To increase our sample and thereby to improve our statistics we combined the results from the PIONIER survey with those from FLUOR, resulting in a total of about 130 stars observed and available for the statistical analysis. The targets for both surveys were selected carefully in order to avoid any relevant selection bias that could affect the statistics.

In addition to the higher efficiency, PIONIER provides a few advantages over FLUOR. The simultaneous use of four telescopes allows the closure phase of the detected systems to be measured, which can only be achieved by combining the light from combinations of three telescopes. This quantity measures the deviation of the brightness distribution of an observed target from point symmetry, allowing a star surrounded by a dust disc and a star orbited by a faint, close companion, which otherwise have a very similar observed visibility deficit, to be distinguished. Another advantage is the fact that PIONIER data can be obtained with a small spectral resolution of three or seven spectral channels across the $H$-band. This information allows one to constrain the spectral slope of the detected excess and thus the emission mechanism and — in case of thermal emission — the temperature of the dust.

### Survey results

From our PIONIER survey we have found that nine out of the 85 targets which proved to be suitable for our analysis (~11 %) show a significant excess, typically around 0.5 to 1 %, above the stellar photosphere. Five targets were found to have a faint, stellar companion and needed to be rejected as exozodis. These latter detections were analysed in a separate study and interesting implications for the formation of binary stars were found (Marion et al., 2014). One more star was rejected from the analysis because it shows hints of post-main sequence evolution which could cause other effects, such as mass loss or strong stellar winds that might mimic an exozodi in our data.

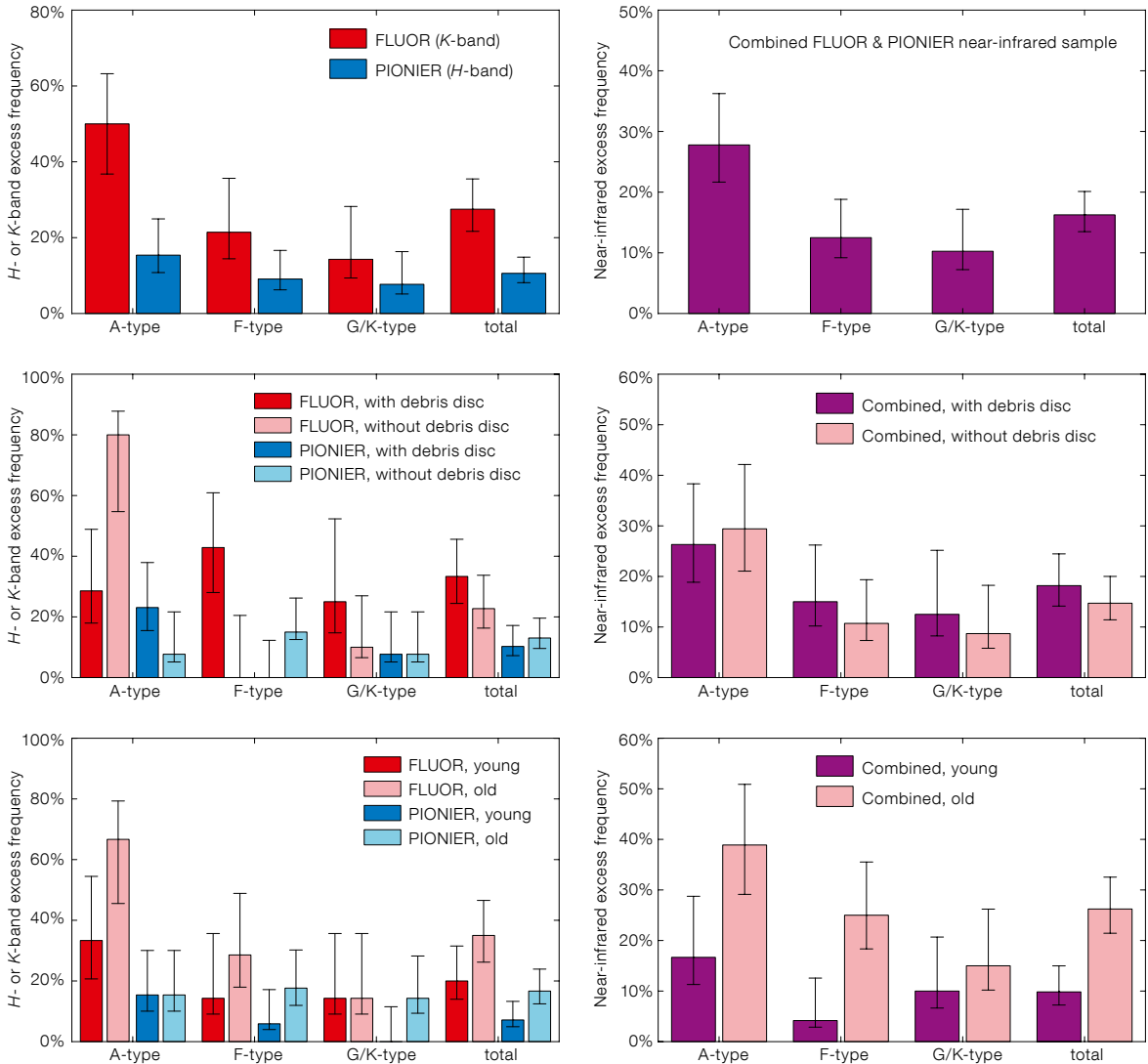

Figure 4. Single (left) and combined (right) statistics from the FLUOR (red) and PIONIER (blue) sample. The top row shows the excess fraction with respect to the stellar spectral type, the middle row shows the same, but, in addition, separated for stars with and without a debris disc detected, and the bottom row the excess fraction for different spectral types and separated for stars younger and older than the median age in each spectral type bin.

A detailed statistical analysis of the combined PIONIER and FLUOR sample (Figure 4) shows that the distribution of the exozodiacal dust detection rate, with respect to the stellar spectral type, is similar to that of debris discs. This suggests that both kinds of dust discs are produced by circumstellar material originating in the protoplanetary disc during the planet formation process. However, there is no correlation between the presence of a debris disc and an exozodi very hot dust emitting significantly in the *H*-band needs to be very close to the star, in the case of scattered light the dust could be slightly more distant, closer to the habitable zone. At that location, the dust has a stronger impact on the detectability of habitable Earth-like planets.

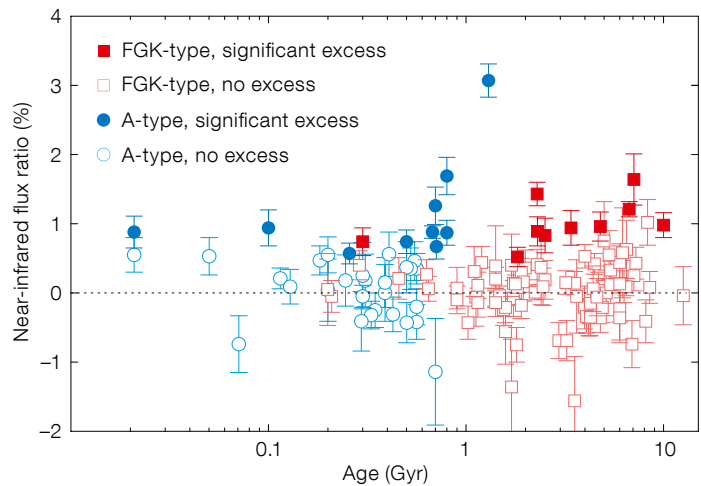

Figure 5. Excess levels for the stars in our sample shown against stellar age. Filled symbols are for stars with significant excess, while empty symbols are for stars without a significant detection. The sample is separated into A-type stars and stars of later spectral types, in order to account for the different main sequence lifetimes of these stars. The star with a 3% flux ratio is α Aql (see Absil et al. [2013] for details).

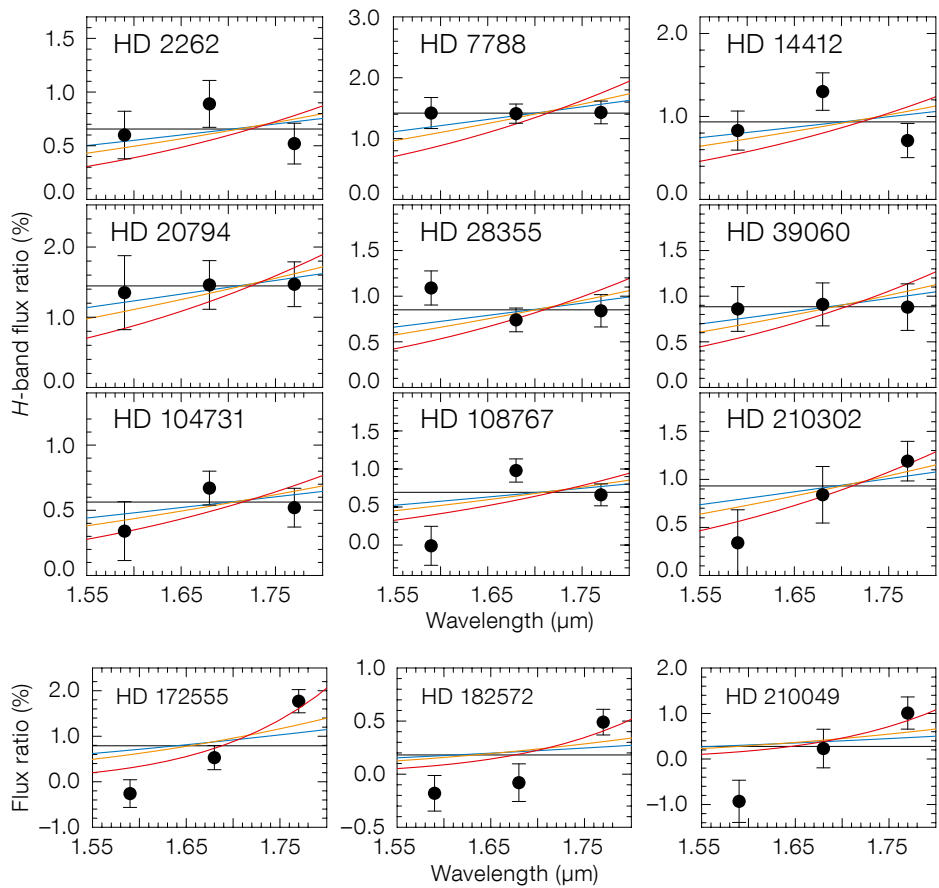

Figure 6. Spectral slopes of the excesses detected with PIONIER (first three rows) and some potential excesses only identified by their spectral slopes, considered tentative (fourth row). The horizontal, black line illustrates the case of purely (grey) scattered stellar light, while the coloured curves show the shapes expected from blackbody thermal emission at different temperatures. Error bars are 1σ.

### Future prospects

Our sample of near-infrared bright exozodis provides an excellent basis for further studies. At the same time, our observations are very timely for the new VLTI instruments, GRAVITY and MATISSE, expected to arrive in 2015 and 2016, respectively. GRAVITY will enable us to observe the discs in the *K*-band, at slightly longer wavelengths than PIONIER, thereby better constraining the slope of the excess and thus the emission mechanism. From the increased emission in *K*-band compared to *H*-band, GRAVITY will also allow the excesses to be measured at very high significance. This will allow detection of small variations in the excess levels, indicative of variation in the mass and location of the dust, enabling strong constraints to be put on its evolution. MATISSE will cover the range of even longer wavelengths up to the *N*-band in the MIR, constraining very well the most relevant region in which the dust emits (Figure 7). Combining all VLTI instruments, we will be able to study in detail the dust distribution and composition in these systems. The strength of combined modelling of different interferometric data has — to some extent — already been shown by our team (Defrère et al., 2011; Lebreton et al., 2013).

One of the main questions to be answered concerns the nature of any connections between the hot dust detected in the NIR, the warm dust in the habitable zone and even the colder dust further out. So far there are indications that there might be an anti-correlation visible in our data. This might suggest that there is indeed no physical correlation between the two phenomena. On the other hand, it is important to note that we are only able to detect the brightest, most extreme exozodis with current instrumentation, and the debris discs detectable so far are also at least a few times more massive than the Kuiper Belt in the Solar System. Thus, a significant fraction of both kinds of discs remain undetected and a correlation might be hidden by that fact.

A very surprising result is that the detection rate (Figure 4) and excess levels (Figure 5) of exozodis do not decrease with the ages of the systems. This would be expected for any phenomenon that evolves over time, such as a belt of colliding planetesimals. In contrast, the exozodis seem to be caused by a stochastic process or a process that can be triggered stochastically at any time during the evolution of the planetary system. One example could be a planetary collision; however, given the high detection rate of exozodis, the short lifetime of the dust and the expected scarcity of such events, this is not a likely explanation for the detected systems. Thus, the independence of the detection rate and excess levels from the ages of the systems surveyed remains enigmatic.

An important result from PIONIER derives from the spectrally dispersed data, enabling us to constrain, for the first time, the spectral slopes of a large number of excess detections (Figure 6). The thermal emission of even the hottest dust is increasing towards wavelengths longer than the *H*-band. A consequence of this thermal emission would be an increasing disc-to-star contrast with wavelength. However, we find for most of our detections that the contrast is rather constant with wavelength, indicating that the emission is dominated by starlight scattered off the dust grains. This unexpected result has important implications for the dust distribution in these systems. While between the presence of NIR- and MIR-detected dust. This might suggest the presence of planets in the systems with MIR bright dust emission that prevents the dust from the outer regions migrating further in, where it would be detected in the NIR. Instead the dust would pile up near the orbits of those planets.

In the context of future observational perspectives, two instruments in the northern hemisphere will soon provide critical complementary information on exozodis: the CHARA/FLUOR instrument in the NIR and the LBTI in the MIR. A new mode at CHARA/FLUOR will provide spectrally dispersed observations of northern objects inaccessible from Paranal, while the LBTI will observe with unprecedented sensitivity a sample of 50 to 60 nearby main sequence stars to characterise the faint end of the exozodi luminosity function (see first LBTI results in Defrère et al. [2015]). However, the efficiency of both instruments is limited due to the use of only one interferometric baseline at a time and the need for a large number of observations to reach the nominal sensitivity. Furthermore, the LBTI exozodi survey is designed for broadband detection of faint levels of dust and will give limited constraints on the dust properties. Only a complete exploitation with both NIR and MIR observations, and in particular those of the near-future instruments at the VLTI, will allow us to precisely characterise exozodiacal dust and study its origin and evolution, even if this is only possible for systems bright enough to be detected by our survey.

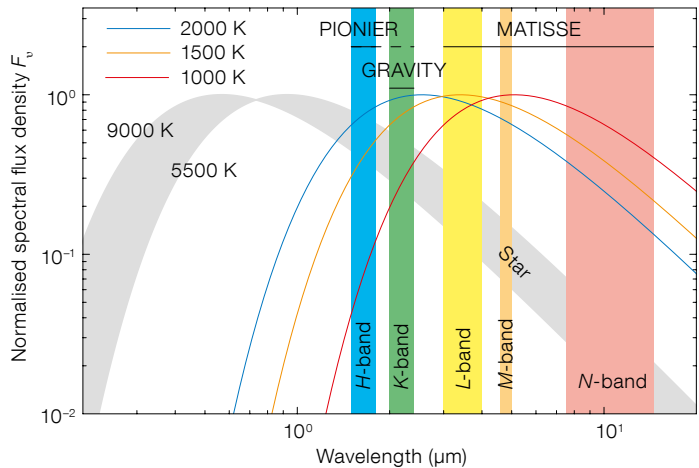

Figure 7. Illustration of the spectral coverage of PIONIER and the future VLTI instruments GRAVITY and MATISSE. The shape of the stellar emission is shown in grey as the blackbody emission of two different, typical stellar temperatures. Dust blackbody curves are shown as coloured lines. The spectral coverage of the VLTI instruments is illustrated by the coloured, vertical bars.

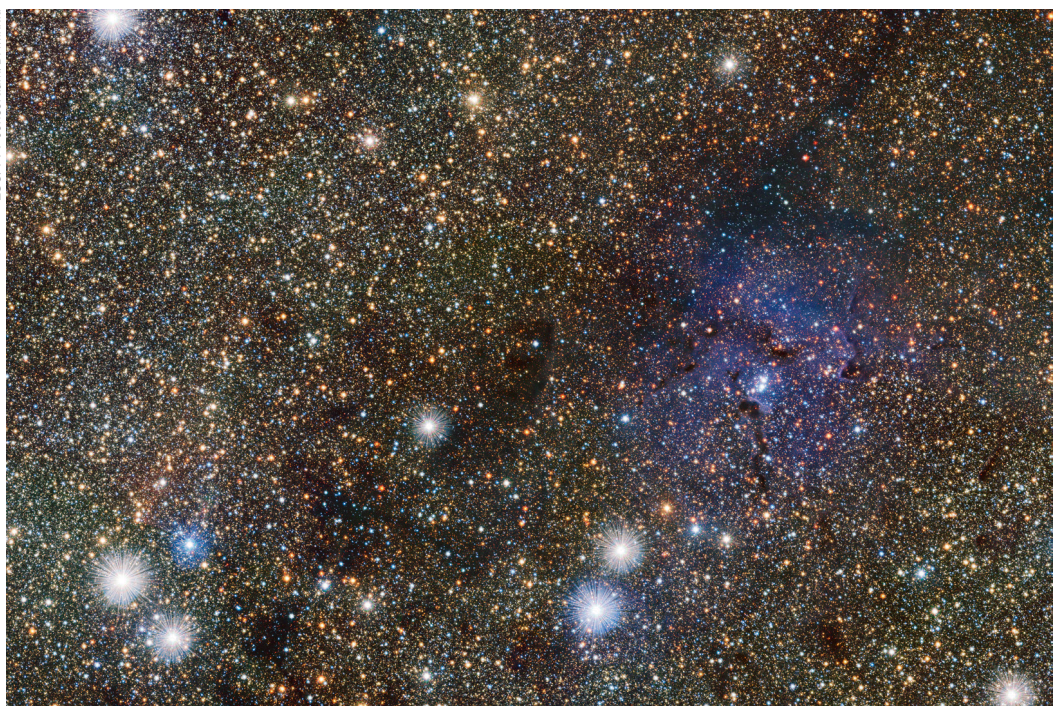

VISTA VVV (the Variables in the *Via Lactea* Public Survey) near-infrared colour image (*JHKs*) of the centre of the Galactic H II region M20 (NGC 6514). At near-infrared wavelengths the line emission is weaker than in the optical and the dust extinction lower, so background stars are easily located. Two Cepheid variables on the other side of the Galactic disc at 11 kpc were detected in this field; see Release eso1504 for details.